# Thermal transport enhancement of hybrid nanocomposites; impact of confined water inside nanoporous silicon

Mykola Isaiev[1*], Xiaorui Wang[2], Konstantinos Termentzidis[2], and David Lacroix[1]

[1]*Université de Lorraine, CNRS, LEMTA, Nancy F-54000, France*

[2]*Univ Lyon, INSA-Lyon, CETHIL CNRS-UMR5008, F-69621, Villeurbanne, France*

*Corresponding author: mykola.isaiev@univ-lorraine.fr*

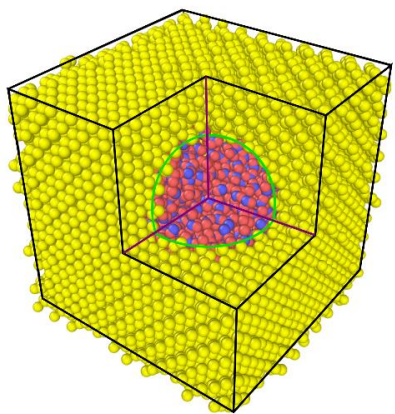

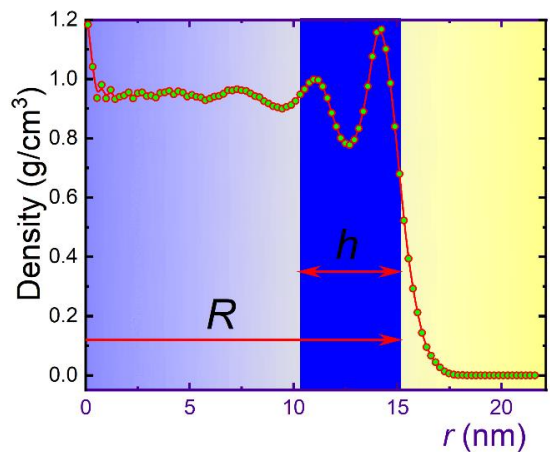

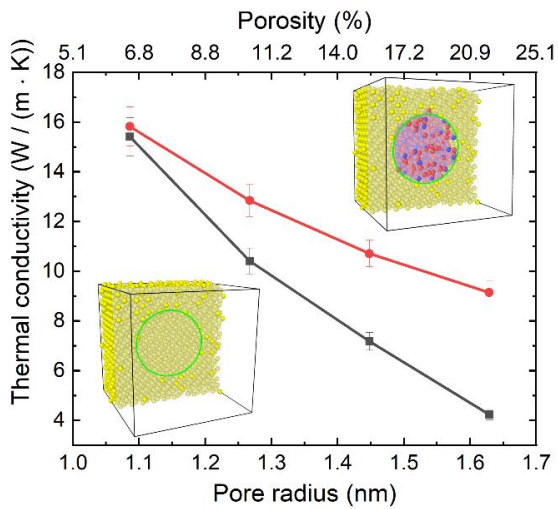

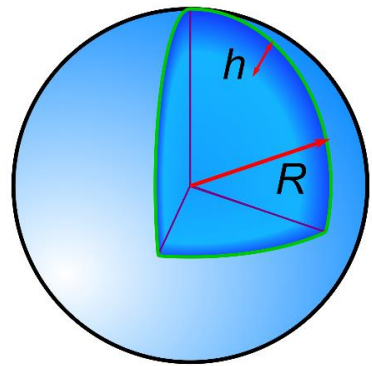

The thermal transport properties of porous silicon and nano-hybrid "porous silicon/water" systems are presented here. The thermal conductivity was evaluated with equilibrium molecular dynamics technique for porous systems made of spherical voids or water-filled cavities. We revealed large thermal conductivity enhancement in the nano-hybrid systems as compared to their dry porous counterparts, which cannot be captured by effective media theory. This rise of thermal conductivity is related to the increases of the specific surface of the liquid/solid interface. We demonstrated that significant difference, more than two folds, of thermal conductivity of pristine porous silicon and "porous silicon liquid – composite" is due to the liquid density fluctuation close to "solid-liquid interface" (layering effect). This effect is getting more important for large specific surface of the interfacial area. Specifically, the enhancement of the effective thermal conductivity is 50 % for specific surface area of 0.3 (1/nm), and it increases further upon the increase of the surface to volume ratio. Our study provides valuable insights into the thermal properties of hybrid liquid/solid nanocomposites and about the importance of confined liquids within nanoporous materials.

Manipulation of heat flux at different scales is one of the key factors for improving reliability and lifetime of different devices (electronics, optics, spintronic, etc.). Typically, efficient thermal management gives the possibility to overcome the occurrence of "hot-spots" in the semiconductor's materials[1], which are the base components of nano-electronics. Therefore, tuning the thermal conductivity ($\kappa$) of materials through nano-structuration, as well as, understanding heat transport in those compounds is crucial for tomorrow practical applications, e.g. 5nm MOFSET technology[2].

From this point of view, nano-hybrids systems like the nanofluids, are well known media that improve heat transport. For instance, one can achieve important enhancement of $\kappa$ of a liquid by adding inside only a small amount of nanoparticles[3]. Such phenomena arise mainly because of significant surface to volume ratio of nanoparticles incorporated in the fluid[4]. This makes nanofluids promising candidates in various cooling applications[5,6]. Nevertheless, nanofluids are liquids, and they are often incompatible with electronic solid-state technologies.

Another important family of materials with significant interfacial exchange area is nanoporous materials. From both experimental[7–9] and theoretical point[10–13] of view, it is well-known that such materials demonstrate significant reduction of thermal conductivity compared to the bulk ones. The main mechanism of the $\kappa$ reduction in nanoporous materials is related to the phonon scattering on the pore edges. Thus, the porous matrix filled by a liquid agent shall exhibit similar behavior as nanofluids. In particular, significant rise of thermal conductivity of porous silicon-liquid composite system while compared to the pristine ones was already observed experimentally[14,15]. However, the nature of heat transport is significantly different for nanofluids and for nano-hybrids systems based on porous materials. In the latter case a substantial part of the thermal energy is transferred through the solid matrix[16]. In crystalline matrix (like for the porous silicon), one should consider phonon contribution to heat transfer, with associated sub-issues like scattering phenomena at the interfaces involving interfacial boundary resistance[8,17,18] and the presence of the adsorbed liquid layer close to the interface[19]. Moreover, strong confinement substantially modifies properties of water[20]. The above-mentioned issues are key points to control heat transfer by using hybrid-like porous composites in nanoscale devices. The understanding of thermal transport mechanisms in nanoporous crystalline matrix-liquid composite is at the core of this article; investigations are carried out using atomistic molecular dynamics (MD) approach.

Ming Hu et al[21] have already reported a significant increase and a nonmonotonic dependence of the overall interfacial thermal conductance between quartz and water layer due to the freezing of water molecules at extremely confined conditions and the vibrational states match between trapped water and the solid. The behaviour of water molecules within nanopores with crystalline pore walls has been examined by Watanabe[22] and it has been found to be very sensitive to the pore structure (shape of the pore section and periodicity) and to the degree of pore filling with water.

In our study the matrix is crystalline silicon with diamond-type lattice with lattice parameter ($a_0$) equals to 5.43 Å. Simulations of solid phase are performed using the Stilling-Weber[23] interatomic potential. The nano-porous silicon (np-Si) systems have porosity in the range 3% < P < 28%. Concerning nanohybrid (nh) configurations, they are like the dry ones the voids are filled with water molecules. The number of water molecules was chosen to reach the averaged density inside pore approximately equals to 1 g/cm³. The SPC/E water model[24] is used for modelling the interactions between water atoms. The interaction forces between oxygen atoms and silicon atoms are treated with the Lenard-Jones potential with parametrization taken from[25]. Two sizes of simulation box are considered with length of $8a_0 \times 8a_0 \times 8a_0$ and $10a_0 \times 10a_0 \times 10a_0$. Eventually, periodic boundary conditions are used in all directions. All simulations were carried with the use of Large-scale Atomic/Molecular Massively Parallel Simulator (LAMMPS)[26].

For $\kappa$ evaluations the equilibrium molecular dynamics (EMD) approach, which is based on the Green-Kubo equation for autocorrelations of heat flux, is used:

$$\kappa_{\alpha\beta} = \frac{1}{V k_b T^2} \int_0^\infty dt \langle J_\alpha(t) \cdot J_\beta(0) \rangle, \qquad (1)$$

where $V$ is the system volume, $k_b$ is the Boltzmann constant, $T$ is the system temperature, $J_\alpha(t)$ is the heat flux in direction $\alpha$ at time step $t$.

Before calculations of heat flux correlations, the systems were equilibrated during 2 ns with the use of canonical ensemble. Snapshots of equilibrated systems for the two box sizes with different pore radii are presented in insets of Fig. 1. After equilibrium, we performed NVE integration during 20 ns while heat fluxes in each direction was recorded. The correlation time length was set to 2 ps, and the sampling interval was 40 ps.

First, the $\kappa$ of a bulk water sample was simulated to test the reliability of our EMD modelling with SPC/E potential. In the present work, the $\kappa$ of pure water was found to be equal to 0.86 W/(m K). This value is higher than the experimental one (0.591-0.607 W/(m K)), but it matches well with other molecular dynamics studies[27] for the same water model.

The thermal transport properties of several np-Si and nh systems with spherical pores are investigated. Thermal conductivities of pristine and nanohybrid systems are presented in the Fig. 1 as a function of pore radius (lower-axis) and porosity (upper-axis) for two sizes of simulation domain. As expected, a strong reduction of ($\kappa$) is observed in dry nanoporous samples as porosity increases. Additionally, Fig. 1a clearly shows significant enhancement of $\kappa$ of nh system as compared to the pristine ones for the simulation domain sizes equal to $8a_0 \times 8a_0 \times 8a_0$. The $\kappa$ rise continuously increases with the pore radius, and it even double for the biggest considered radius ($R = 3a_0 = 1.63$ nm). Fig. 1b depicts the results when the simulation domain has length size equals to $10a$. The relative increase of nh $\kappa$ is less pronounced even for the same porosities as those computed for the $8a \times 8a \times 8a$ box size (e.g. the penultimate point in Fig. 1b and the last one in Fig. 1a). Since the porosity is a volumetric characteristic parameter, one can conclude that the $\kappa$ enhancement between dry and wet systems has a nanoscale nature where surface phenomena dominate. This issue is further investigated to determine the relevant parameters that rule heat transport. The dependence of the $\kappa$ for both considered box sizes on the specific surface area of a pore is presented in Supplementary Materials S1.

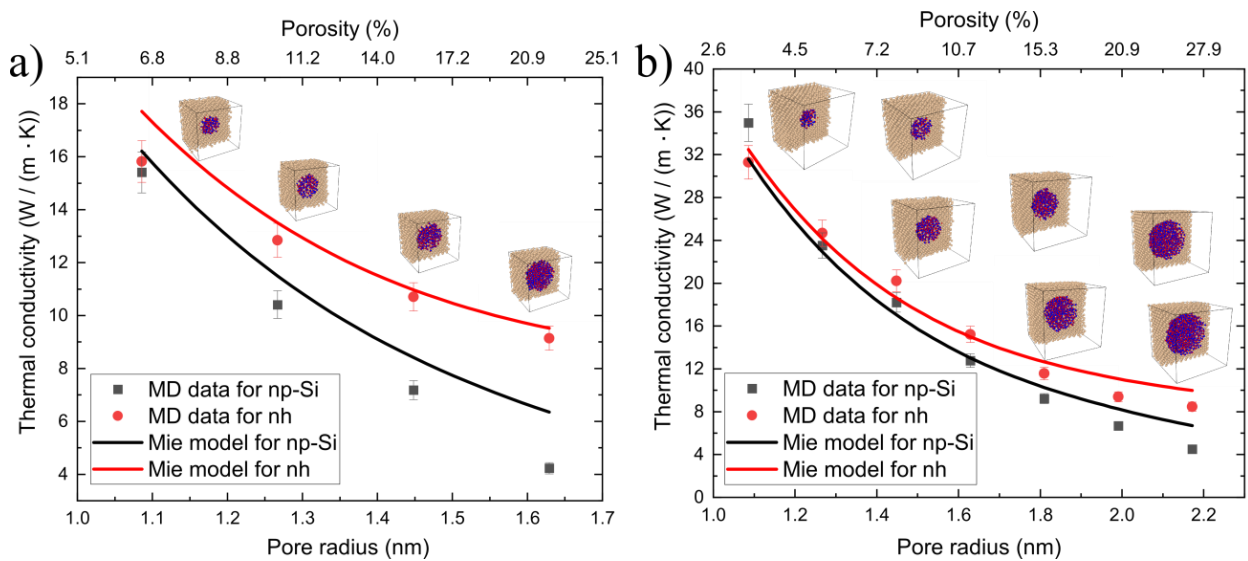

Figure 1. Dependence of $\kappa$ on the pore radius (lower axis) and porosity (upper) axis for the two domains' sizes: $8a_0 \times 8a_0 \times 8a_0$ (a) and $10a_0 \times 10a_0 \times 10a_0$ (b). The black points correspond to the dry samples, while the red ones to the wet samples. The lines are results of the modelling based on the Mie scattering theory. The insets show MD snapshots for corresponding pore' radius (porosity).

The dependence of the $\kappa$ as a function of a pore specific surface area ($\xi = A_{por}/V_d$, where $A_{por}$ is the pore surface area, $V_d$ is the volume of simulation box) is presented in the Fig. 2 for both box radii. The magnitude of the thermal enhancement significantly increases upon the specific surface area.

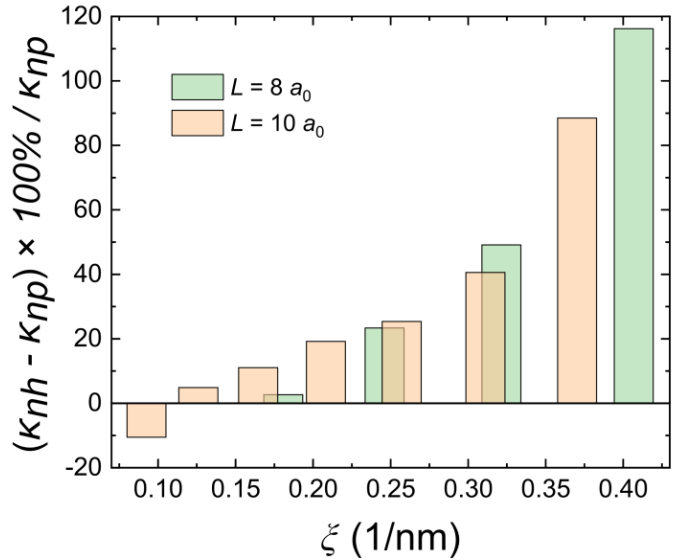

Fig. 2. Thermal conductivity enhancement due to nanoconfined water in nanocomposite as a function of the specific surface area, $\kappa_{nh}$ is the thermal conductivity of the nanohybrid, and $\kappa_{np}$ is the thermal conductivity of the dry nanoporous silicon

According to Nan et al [28] effective medium theory, $\kappa$ of a composite can be modelled as follows:

$$\kappa_{comp} = \kappa_m \times F, \qquad (2)$$

where $\kappa_m$ is the matrix thermal conductivity, $F$ is the factor which describes the effect of material constitution. A simple description of the volumetric factor can be taken from two phase Maxwell effective model [29], it reads

$$F = \frac{\kappa_f(1 + 2\alpha) + 2\kappa_m + 2P(\kappa_f(1 - \alpha) - \kappa_m)}{\kappa_f(1 + 2\alpha) + 2\kappa_m - P(\kappa_f(1 - \alpha) - \kappa_m)}, \qquad (3)$$

where $\kappa_f$ is the filler TC (fluid in our case), $P$ is the porosity, $\alpha = \delta/R$ is the dimensionless parameter which depends on the Kapitza length ($\delta = R_k \kappa_{bulk}$, where $R_k$ is the thermal boundary resistance).

It should be noted, that actual $\kappa_m$ is lower compared to the thermal conductivity of a bulk material ($\kappa_{bulk}$), because of the phonon scattering on a pore's edge. Specifically, when phonon mean free path ($l_{mfp} \sim 300$ nm in bulk silicon[30]) is comparable or bigger than inclusion characteristic size, the matrix thermal conductivity reads as follows according to Minnich and Chen model[31]

$$\kappa_m = \frac{\kappa_{bulk}}{1 + \xi \cdot l_{mfp}/4}, \qquad (4)$$

Here for $\kappa_{bulk}$ we use value evaluated by EMD for bulk silicon at room temperature with the S-W potential [32]. $l_{mfp}$ is the phonon mean free path in a bulk material, $\xi$ is the specific surface area, and it can be considered as an inverse characteristic length. The mean free path was estimated based on the fitting of the treated MD results with the Eq. (4) (see Supplementary Materials S2 for details). With this fit we have estimated that the mean free path (mfp) equals to $l_{mfp}$ = 460±20 nm for bulk silicon, the latter value is in the same order of magnitude as the one given in the literature ($l_{mfp}$ = 300 nm)[33,34].

It should be noted that the mentioned above model assumes geometrical regime of scattering [35]. It works quite well when the pore radius is much bigger than phonon wavelength (size parameter $\chi \gg 1$, $\chi = qR$, $q$ is the wavevector). In our case, the mean phonon wavelength is in the same range as pore radius size. Therefore the previous model should be modified, for instance with the use of Mie scattering theory.

For this purpose, one needs to modify the scattering cross-sectional area used in the model of Minnich and Chen[31]. In their work, it is assumed that it equals to the projected area of a pore ($\sigma_{geom} = n\pi R^2$, $n$ being the number density of scattering centers). However, when wavelength of the phonons is in the same order as a pore diameter, Mie theory[36] gives the following expression for the cross-sectional area

$$\sigma_{Mie} = \sigma_{geom} \times Q_{sca}, \qquad (5)$$

Where

$$Q_{sca} = \frac{2}{\chi^2} \sum_{n=1}^{\infty} (2n+1)\,(|a_n|^2 + |b_n|^2), \qquad (6)$$

$$a_n = \frac{\psi_n'(m\chi)\psi_n(\chi) - m\psi_n(m\chi)\psi_n'(\chi)}{\psi_n'(m\chi)\zeta_n(\chi) - m\psi_n(m\chi)\zeta_n'(\chi)} \qquad (7)$$

$$b_n = \frac{m\psi_n'(m\chi)\psi_n(\chi) - \psi_n(m\chi)\psi_n'(\chi)}{m\psi_n'(m\chi)\zeta_n(\chi) - \psi_n(m\chi)\zeta_n'(\chi)}, \qquad (8)$$

$m = v_f \rho_f / (v_m \rho_m)$ is the ratio of acoustic impendence of a fluid filler and a matrix ($v$ is sound velocity, $\rho$ is density), $\psi_n(\chi)$ and $\zeta_n(\chi)$ are Riccati-Bessel functions.

As a result, equation (Eq.4) can be corrected considering Mie scattering dependence as follows: $\kappa_m = \frac{\kappa_{bulk}}{1 + \delta \cdot \xi \cdot l_{mfp}/4}$, where $2\delta = Q_{sca}$. Supplementary Materials S2 presents comparisons between np-Si thermal conductivities obtained with MD simulations and their counterparts derived from the two above discussed effective medium models: Minnich and Chen[31] and Mie scattering. For the Mie approach the recalculated mfp, with the use of mean square minimization method, was $l_{mfp}$ =450±20 nm. The simulation results of the Mie model are presented in the Fig. 1.

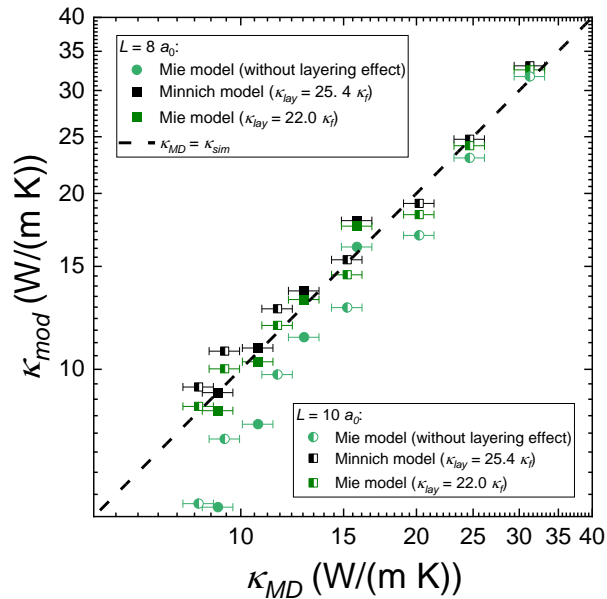

Fig. 3. The diagram presents the correlation between MD and analytical results for pristine porous silicon for nanohybrid system. Filled squares correspond to the simulation domain length equal $8a_0$, half-filled – $10a_0$. Black color represents results obtained based on the Minnich and Chen model, and olive color shows results of Mie scattering model. Squares and circles are devoted to the situations with and without layering effect for the nanohybrid system, respectively.

In the Fig. 3 the correlations between the MD simulations and the analytical modeling are plotted. As one can see from the figure, the direct use of Eq. (4) (data drawn as circles in Fig. 3) gives significantly underestimated $\kappa$ values while compared to the MD ones. The difference shall arise because of the manifestation of the water layering effect, i.e. the presence of a thin adsorbed layer of liquid with higher density close to solid/liquid interface[37]. This layering effect was observed experimentally[38]. In nanofluids the enhancement of $\kappa$ is also related to this phenomena[39].

The occurrence of a high-density layer close to the interface is revealed by the presence of the water density oscillations (Fig. 4). For evaluation of its thickness we used an approach which is based on the equimolar definition of separated surface[40] (see Fig. 4). More specifically, the outer interface was defined as the surface across the radius, which corresponds to half height of the density profile $((\rho_1 + \rho_2)/2$, where $\rho_1 = 0$ g/cm³ is the density of water outside the sphere and $\rho_2$ is maximum density). For the inner surface we chose a sphere across the density, which corresponds to the average value between density of the second maximum ($\rho_3$) and minimum ($\rho_4$). We estimate the thickness of this layer ($h$) as the thickness of the first two peaks, it was found to be equal 0.5 nm. This value is slightly bigger than those evaluated previously from the Gibbs adsorption (0.4 nm)[41] because of the nature of the curved interface. Computed water density profiles for several pore size is reported in Supplementary Material S3.

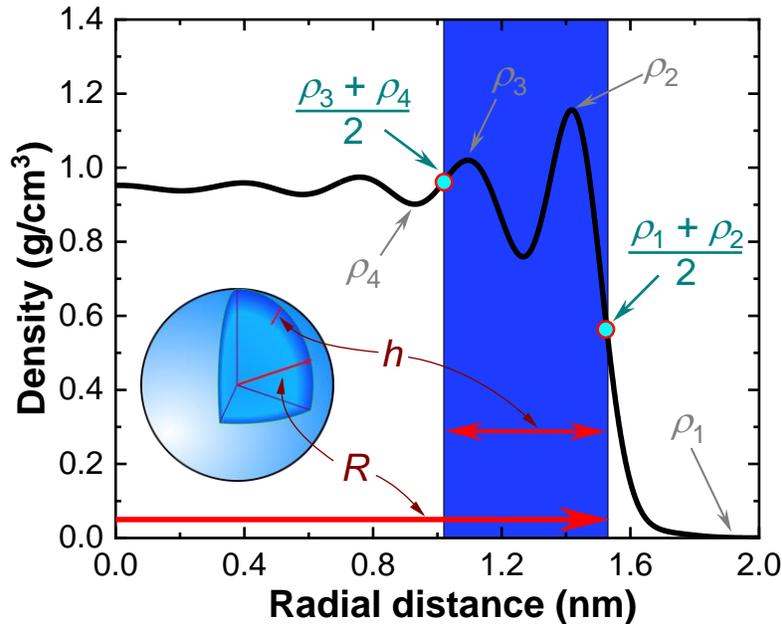

Figure 4. Sketch of the water density profile variations inside a pore. The chosen volume of the adsorbed layer is presented by blue color. The inset demonstrates schematic representation of the structure of confined water.

To consider the impact of this layer on the system' $\kappa$, we adopted a model that it is typically used for the estimation of $\kappa$ in nanofluids[39]. In frames of this model, thermal conductivity of a liquid ($\kappa_f$) in Eq. (3) should be modified as follows

$$\kappa_{f_{mod}} = \kappa_f \frac{(2(1-\beta) + (1+\gamma)^3(1+2\beta))\beta}{-(1-\beta) + (1+\gamma)^3(1+2\beta)},$$
(9)

where $\beta = \kappa_{lay}/\kappa_f$ and $\gamma = h/(R-h)$, $\kappa_{layer}$ is the adsorbed layer thermal conductivity, and $h$ is it thickness (see Fig. 6).

And the resulting equation for the volume factor is

$$F(P) = \frac{\kappa_{f_{mod}} + 2\kappa_{bulk} + 2P(\kappa_{f_{mod}} - \kappa_{bulk})}{\kappa_{f_{mod}} + 2\kappa_{bulk} - P(\kappa_{f_{mod}} - \kappa_{bulk})}.$$
(10)

Here, there is no extra parameter such as interfacial boundary resistance between solid and liquid ; all the physics is set in absorber layer model.

Squares in Fig.3 corresponds to model values that consider the layer effect (Eq. 10). We vary the value of $\kappa$ of a boundary layer to fit these two sets. As a criterion of the fitting, the minimum of mean least deviation was used. We found the following values of the boundary layer: $\kappa_{lay} = 25.4\kappa_f$ for the Minnich model and $\kappa_{lay} = 22.0\kappa_f$ for the model based on the Mie scattering. The latter is higher than typical $\kappa$ for nanofluids (for example $\kappa_{lay} = 10\kappa_f$ in Yu and Choi[39]). The difference in the sign of the interfacial curvature, in the case of a liquid covering nanoparticle (nanofluid) or in the case of a liquid inside porous matrix (nanocomposite), might explain observed differences between the nanofluids and the confined water inside a pore. Similar curvature dependence was found by K. Falk et al[42] for the interfacial friction of water at graphitic interfaces with various topologies. They have shown that the friction coefficient exhibits a strong curvature dependence, driving to improved transport of water in nanometric carbon nanotube membranes. The enhanced $\kappa_{lay}$ observed here compared to the one of nanofluids might have the same nature as the lower friction coefficient for the negative (nano-hybrid composite) or positive (nanofluids) curvature.

To summarize, in this work, we considered features of thermal transport in a nano-hybrid composite system with the use of MD and analytical modelling. Significant enhancement of the effective thermal conductivity in the nanocomposite system compared to the pristine one was observed (see Fig. 2). This enhancement is getting more pronounced with the increase of the size of the pores. Indeed, TC is doubled in nanocomposite system compared to the pristine porous silicon for the biggest considered specific surface area. For the analytical modelling, the approach based on Minnich and Chen model[31] was used. Yet, as considered pore sizes and phonons wavelength were in the same order of magnitude, the above-mentioned approach was modified based on the Mie scattering theory. Furthermore, the development of such methodology was improved to consider the presence of the surface adsorbed layer of liquid with higher density. It was shown an excellent correlation between the MD data and results of analytical modelling. Eventually, the thermal conductivity of the adsorbed layer was estimated to be 25 times greater than in the bulk water. This estimation is based on this correlation and minimization of the MD outputs.

## Data Availability

The data that support the findings of this study are available from the corresponding author upon reasonable request.

## Acknowledgment

The publication contains the results obtained in the frames of the project "Hotline" ANR-19-CE09-0003. This work has been partially funded by the CNRS Energy unit (PEPS Cellule ENERGIE 2019) through the project "ImHESurNaASA". M. Isaiev and D. Lacroix want also to acknowledge the partial financial support of the scientific pole EMPP of University of Lorraine. The authors thank to the mesocenter EXPLOR of the University of Lorraine and National Computing Center for Higher Education (eDari project A0080907186) for the providing of computational facilities. Finally, authors want to acknowledge V. Lysenko from Institut Lumière Matière (Lyon) for fruitful discussions.

Supplementary materials.

### S1 Dependence of the thermal conductivity on the specific surface area

The Fig. S1.1 presents dependence of the thermal conductivity calculated by molecular dynamics as a function of the specific surface area for both box sizes. Additionally, in the inset of the Fig. S1.1 the $\kappa_m = \kappa_{MD}(2 + P)/(2 - P)$ for pristine np-Si as a function of specific surface area by dots is depicted. The solid line corresponds to the fitting of the MD points with Eq. (4). With this fit we estimated that the mean free path (mfp) equals to $l_{mfp}$ = 460±20 nm for bulk silicon, the latter value is in the same order of magnitude as the one given in the literature ($l_{mfp}$ = 300 nm).

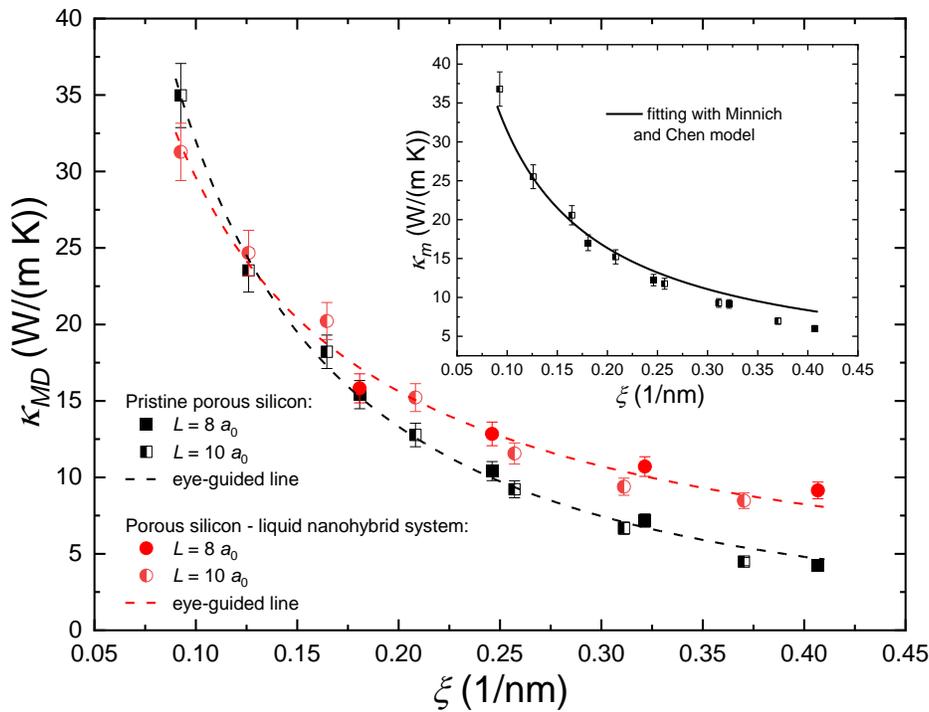

Figure S1.1. Thermal conductivity of a pristine and a nanocomposite porous silicon as a function of the inverse characteristic length $\xi$. The results for the dry systems are with black symbols, while for the wet systems with red. Full color points are for the small domain simulations and half colored for the large one. Inset: thermal conductivity normalized by $F$ (Eq. 3) for dry samples, the solid line gives the fitting obtained with Minnich and Chen model[31]

Considering the inset of Fig. S1.1, it can be noticed that Minnich and Chen model[31] fits MD results well for small inverse characteristic length values ($\xi$).

## S2 Correlations of Minnich and Chen Model and Mie theory-based model with MD results for np-Si

Fig. S2.1 presents comparisons between np-Si thermal conductivities obtained with MD simulations and their counterparts derived from the two discussed effective medium models: Minnich and Chen[31] and Mie scattering. As one can see from the figure, the use of scattering cross-section calculated with the Mie theory gives relatively better correlation with molecular dynamics. This is logical considering dimensionless parameter $Q_{sca}$ (presented in the inset of Fig. S2.1) which exhibits a significant deviation from a constant value in the considered range of pore' radii.

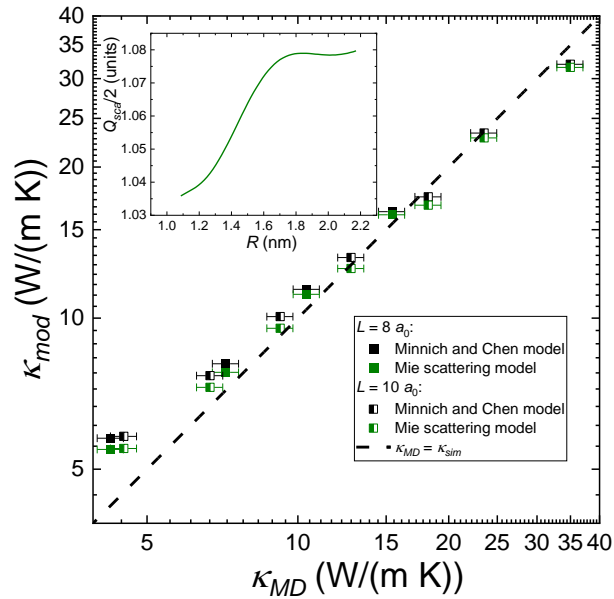

Fig. S2.1. The diagram presents the correlation between MD and analytical results for pristine porous silicon. Filled squares correspond to the simulation domain length equal $8a_0$, half-filled – $10a_0$. Black color represents results obtained based on the Minnich and Chen model, and olive color shows results of Mie scattering model. The inset: $Q_{sca}$ as a function of the pore radius.

**S3 Density profile inside pores.**

The dependence of the water density ($\rho$) on the radial distance from the pore' center ($r$) for different pore radii ($R$) is depicted in figure S3.1. Density profiles inside a pore were calculated by dividing the space by spherical bins with the centers situated in the middle of the pore (and box). The thickness of the shells was equal to 0.02172 nm. The density in each shell was calculated by averaging the water density during 10 ns. As one can see from the density profiles inside the porous matrix, there is a thin layer of adsorbed liquid with higher density.

It should be stated, that for different pore' radius, the density profile has the same structure with the presence of well-define maximum peak and further oscillations. The width of interfacial area is almost the same for all radii as it is depicted in the inset of the figure, where the density profiles are shifted by $r_{\rho_{max}/2}$ to align them to the solid/liquid interface, where the water density is equal to the half density of the first peak ($\rho_{max}/2$).

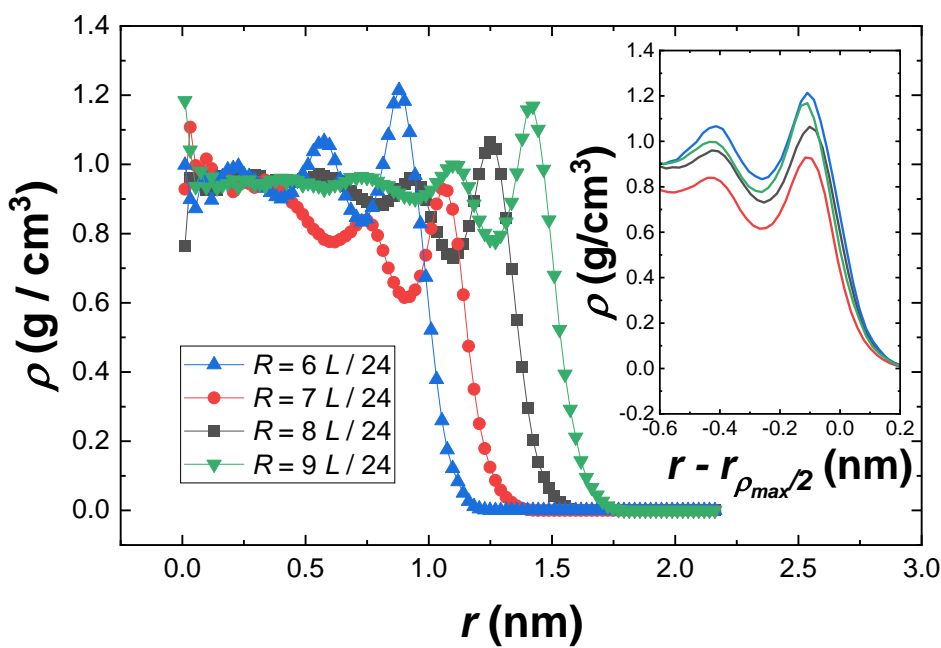

Fig. S3.1 Density distribution of water inside a pore for different pores radii. The inset presents the same density profiles aligned to zero density to stress the effect of the liquid/solid interface on the water density oscillations